# Beyond ITER: Neutral beams for DEMO[a)]

R. McAdams[b)]

*EURATOM/CCFE Association, Culham Science Centre, Abingdon, Oxfordshire OX14 3DB, UK*

In the development of magnetically confined fusion as an economically sustainable power source, ITER is currently under construction. Beyond ITER is the DEMO programme in which the physics and engineering aspects of a future fusion power plant will be demonstrated. DEMO will produce net electrical power. The DEMO programme will be outlined and the role of neutral beams for heating and current drive will be described. In particular, the importance of the efficiency of neutral beam systems in terms of injected neutral beam power compared to wallplug power will be discussed. Options for improving this efficiency including advanced neutralisers and energy recovery are discussed.

## I. INTRODUCTION

On the route to sustainable power from magnetic confinement fusion, the International Tokamak Experimental Reactor (ITER) is currently under construction at Cadarache in France. ITER will operate to produce net output of fusion power that exceeds the heating power by a factor of Q=10 and produce a burning (self-sustaining) plasma for several hundred seconds. However ITER is still an experimental device and will not produce any electricity. Beyond ITER is the DEMO machine which will produce electricity and demonstrate the requisite technologies to allow commercial production of electrical power. Figure 1 shows a route to fusion power from today's tokamaks such as JET, moving through ITER and DEMO to a commercial fusion reactor.

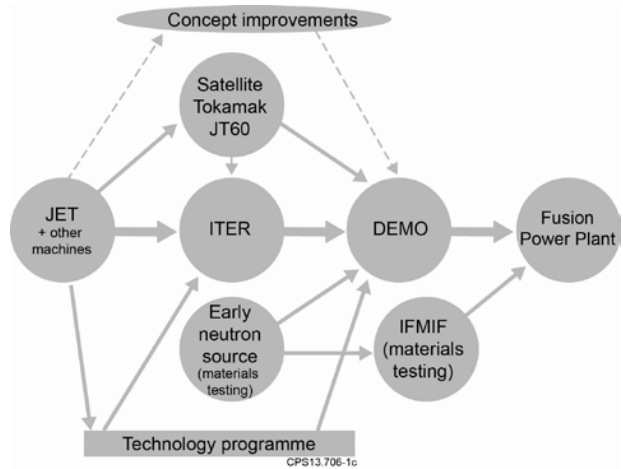

FIG. 1. The route to fusion power

DEMO will link a fusion source with electricity generation and will be the last machine before a commercial fusion reactor. The European Fusion Roadmap[1] calls for construction of DEMO to commence in 2030 at the point where ITER has successfully demonstrated the Q=10 performance. The European programme is currently considering two options[2] as of July 2013 for DEMO: a "pulsed" and a "steady state" option (1A and 2). The main features of these are given in Table 1 below and are compared to ITER. The neutral beam power is required to heat the plasma to reach the burn stage and sustain the pulse length by current drive.

TABLE I. Comparison between ITER and the DEMO options.

|  | ITER | DEMO 1A | DEMO 2 |
|---|---|---|---|
| Major/minor radius (m) | 6.2/2 | 9/2.49 | 8.1/3 |
| Toroidal field (T) | 5.3 | 6.5 | 5.0 |
| Plasma current (MA) | 15 | 16.8 | 19.9 |
| Average electron density ($\times 10^{19} m^{-3}$) | 10 | 9.3 | 7.7 |
| Average electron temperature (keV) | 9 | 13 | 15.5 |
| Pulse length (hrs) | 1 | 2 | 300 |
| Net electrical power (MW) | - | 500 | 500 |
| Neutral beam power MW) Heat/sustain pulse length | 33 | ~100/50 | ~135 |
| Neutral beam energy (MeV) | 1 | 1 | 1 |

This paper will discuss the physics and technological challenges for the neutral beam systems to meet the requirements for DEMO and beyond.

## II. NEUTRAL BEAM SYSTEM REQUIREMENTS FOR ITER, DEMO AND BEYOND

### A. The ITER Neutral Beam Injectors

The ITER neutral beam injectors[3] consists of two beamlines each with an accelerated beam of 40A of $D^-$ ions at 1MeV. This beam passes through a gas neutraliser which gives ~58% neutralisation. Accounting for

---

transmission and other losses the injected neutral beam power for each beamline is 16.7MW. Test stands are currently operating or are under construction in a phased programme to develop the injectors: a half scale ion source at IPP Garching[4] (ELISE - 20A, 60keV, D[-]) and at RFX in Padua[5] a full sized ion source (SPIDER - 60A, 100keV H[-]) and a full sized injector (MITICA - 40A, 1MeV D[-], 16.7MW D[0]). In the ITER injector[6] there is ~0.9MW of backstreaming positive ions, ~9MW of power striking accelerator grids and ~0.7MW of electrons exiting the accelerator.

### B. Choice of beam energy for DEMO

The neutral beam systems for the DEMO options, as given in Table 1, are to heat the plasma to the burn phase and to drive current to meet the pulse length requirements. The choice of beam energy for the DEMO machine is primarily dependent on the current drive efficiency and the shinethrough of the beam onto the machine wall. As can be seen in Table 1 the present beam energy choice is for 1MeV. Studies of the current drive in DEMO[7], have shown that for peaked and flat plasma distributions the current drive efficiency and shinethrough power levels are fairly similar for beam energies of 1 and 1.5MeV.

This choice of energy will allow the experience from design, construction and operation of the ITER injectors to be used directly. A choice of beam energy at 1.5MeV would pose additional technological challenges in terms of power loadings in the accelerator and the ion source backplate together with HV insulation both within the accelerator and in relation to voltage hold off of the ions source from the vacuum wall.

### C. System efficiency requirements for DEMO and future power plants

Commercial viability of fusion power plants depends on minimising the recirculating power used to operate the reactor. This recirculating power can be high. For example, a power plant study[8] shows that producing fusion power of 2.4GW with a net power of 1.56GW, one third of this net power to is required to operate the reactor. Most of this recirculating power (~300MW) is used by the heating and current drive systems.

For neutral beam systems, the power requirement for the neutral beam is given by

$$P_{NB} = \frac{P_{CD}}{\eta_{conv}\eta_{coup}\gamma_{CD}} \quad (1)$$

where $P_{CD}$ is the neutral beam current drive power in the tokamak, $\eta_{conv}$ is the efficiency of electrical power conversion into neutral beam power injected into the tokamak i.e the wall plug efficiency, $\eta_{coup}$ (~1) is the efficiency of coupling the injected power to the plasma and $\gamma_{CD}$ is the current drive efficiency (current driven per unit power coupled into the plasma). This shows the premium to be gained by maximising the wallplug efficiency. The wall plug efficiency for present neutral beam systems[8] is of the order of 20-30% and systems studies[9] show that the required product of $\eta_{conv}\gamma_{CD}$ should be > ~0.25. Thus with current drive efficiencies[10] of ~0.45-0.6 then the neutral beam wall plug efficiency requires a step increase from present systems to at least 0.4-0.55

## III. PHYSICS AND TECHNOLOGY CHALLENGES FOR SYSTEM EFFICIENCY IMPROVEMENT

The potential for improvement in system efficiency lies in a number of areas: improvement in transmission by reduction of the beam halo and core divergence, increased neutralisation efficiency and improved electrical efficiency through the use of energy recovery.

### A. Beam halo and divergence

There is known to be a halo associated with the beam extracted from caesiated H[-]/D[-] sources[11] compared to uncaesiated sources. The situation is shown schematically in Figure 2.

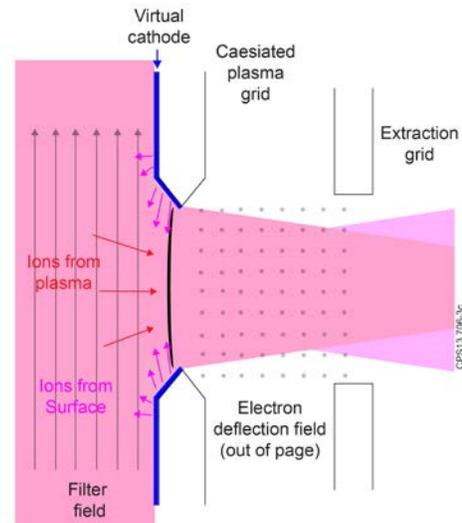

FIG. 2. (Color online). Beam formation in a caesiated negative ion source

Negative ions created by atoms on the caesiated surface are transported across the virtual cathode[12] and sheath into the plasma. Those from the bulk plasma enter the meniscus to form a well ordered beam. Those entering from the surface have high transverse velocities and give rise to the beam halo. This situation has been modelled in both 2D[13] and 3D[14,15] where there is high penetration of the accelerator field into the ion source. Caesium migration onto the accelerator grids may give rise to highly aberrated negative ions due to atoms or backstreaming ions. A search for a suitable replacement for caesium has not been successful so far. There is a deflection in the beam due to the magnetic fields and space charge repulsion of the beamlets and this also contributes to the divergence. These effects can be overcome by off-set aperture steering and modification of the electric field at the rear of the grids[16,17]. This has allowed the performance of the accelerator to become improved as 1MeV is approached.

### B. Improved neutralisation

The gas neutralisation efficiency for D[-] ions at high energies is constant at ~ 0.58. Improvements to this neutralisation will be reflected in the overall system efficiency. Two such methods are discussed.

*1. Photo-detachment neutraliser*

Photo-detachment offers the possibility of attaining high (>90%) neutralisation efficiencies whereby a laser is used to neutralise the ions as they pass through the radiation field. The basic scheme is shown in Figure 3 and various proposals have been made[18,19,20,21,22].

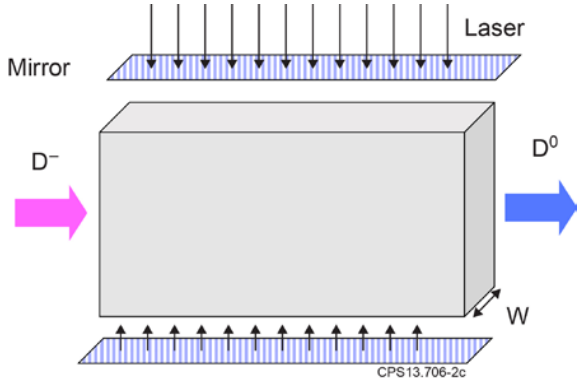

FIG. 3. (Color online). Basic scheme for a photo-detachment neutraliser

The laser power, P, required for a degree of neutralisation f is

$$P = -\frac{hc}{\lambda}\frac{\ln(1-f)}{\sigma}v_b\frac{w}{G} \qquad (2)$$

where $v_b$ is the beam velocity, h is Planck's constant, c is the speed of light, $\lambda$ is the wavelength, $\sigma$ is the photo-detachment cross-section ($3.4 \times 10^{-21}$ m$^2$ at 1064nm), w is the average beam width and G is the gain or number of laser passes through the beam. For w=0.25m and G=100 and a neutralisation efficiency of 95% the laser power requirement is 0.8MW for a 1MeV beam and 1MW for a 1.5MeV beam. Even though the laser efficiency may be <25% the gain in efficiency from the additional neutralisation offsets the power requirement.

The photo-neutraliser has the additional advantage of reducing the gas requirement and will reduce stripping in the accelerator. However extensive development work is required in high power lasers, high reflectivity mirrors, cooling, stability and radiation damage.

*2. Plasma neutraliser*

Ionisation of the gas in the neutraliser would allow advantage to be taken of the much higher cross-sections for neutralisation of negative ions by positive ions and electrons compared to gas molecules. It has the further advantage of lowering the target thickness and thus the gas requirement. This is the basis of the plasma neutraliser[23]. The maximum neutralisation is shown as a function of the degree of ionization in Figure 4 for 1MeV and 1.5MeV. This maximum neutralisation is independent of energy as it depends on a ratio of cross-sections which scale in the same way at these energies. Figure 5 shows the optimum target thickness corresponding to the maximum neutralisation. This is dependent on energy since it scales according to the absolute values of cross-sections. Some development and testing of plasma neutralisers has been done. A magnetic multipole system with microwave heating of the plasma has been proposed and is under development[24,25]. In argon an ionisation degree of 10-25% has been reported. An arc driven plasma operating in hydrogen has also been tested with a 200keV H$^-$ beam[26]. The neutralisation measurements are consistent with a 10% degree of ionization.

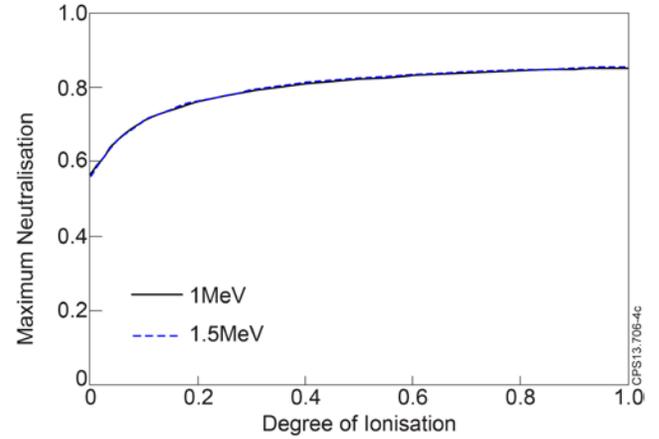

FIG. 4. (Color online). The maximum neutralisation for a plasma neutraliser as a function of the degree of ionisation

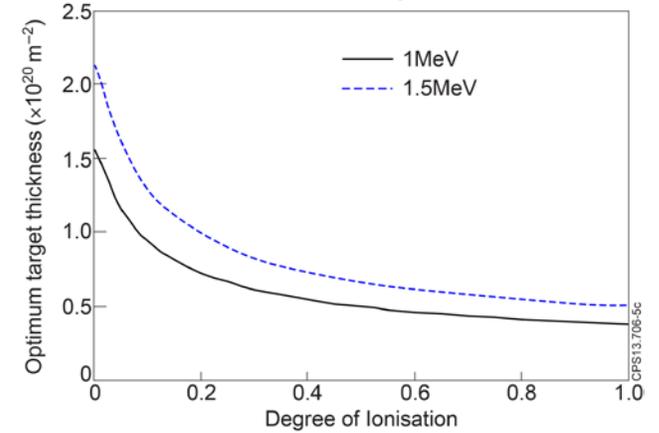

FIG. 5. (Color online). The optimum target thickness for a plasma neutraliser as a function of the degree of ionisation

The ITER gas neutraliser has a degree of ionization of ~0.001% due ionization by the beam (D$^-$,D$^0$,D$^+$) giving rise to an electron distribution with average energy of ~62eV causing further ionisation and stripped electrons at 272eV. It has been proposed[27] that by adding multipole confinement alone that a relatively high degree of ionisation may be achievable. Figure 6 shows the results for the ITER beamline with a 3x1.7x0.4m$^3$ neutraliser with multipole confinement at a cusp separation of 0.18m.

The general development issues associated with the plasma neutraliser relate to achieving relatively high degree of ionisation, the effect of the multipole field on beam divergence and end losses from plasma leaking out of the neutraliser. In the case of the beam driven neutraliser, this has not been tested and also requires the use of high cusp strengths to achieve the ionisation required.

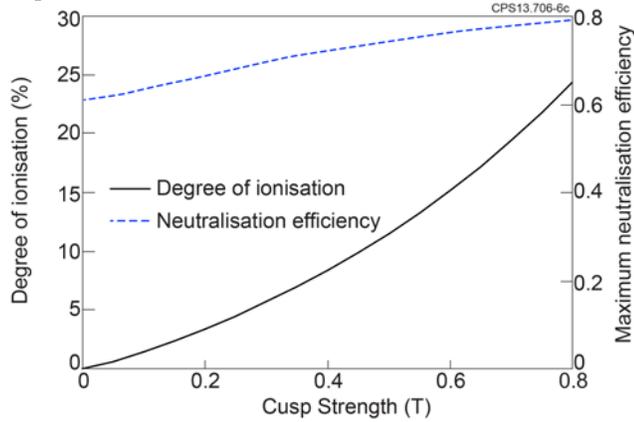

FIG. 6. (Color online). The degree of ionisation and neutralisation efficiency for a beam driven plasma neutraliser

## C. Improved electrical efficiency

The energy efficiency of the injection system can be improved by recirculating the residual negative ions. This is referred to as energy recovery. It has been proposed and tested for positive ions and for negative ions[28-35]. The basic scheme is shown for a negative ion system in Figure 7.

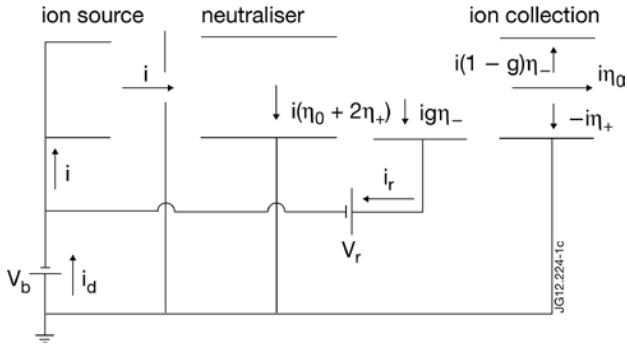

FIG. 7. Energy recovery of negative ions

A negative ion current, i, is extracted and accelerated. After the neutraliser the beam fractions of negative ions, positive ions and neutrals are $\eta_-$, $\eta_+$ and $\eta_0$. In the neutraliser a current of electrons flows to ground. A fraction, g, of the negative ions are decelerated by the power supply at voltage $V_r$ and recirculated. The neutrals are injected into the tokamak and positive ions and remaining negative ions are collected at ground. The high voltage power supply drain current is $i_d = (1-g\eta_-)$. For the ITER case with $\eta_- \sim 0.21$ and 90% recirculation the power in the HV power supply is reduced by 19%. A figure of merit for energy recovery, $D_r^0$, defined as the neutral beam power divided by the HV power, $D_r^0 = \eta_0/(1-g\eta_-)$, is independent of energy at energies > 1MeV as the relevant cross-sections scale in the same way with energy. Since the residual positive ions are created at ground potential in the neutraliser, there is no possibility of recirculating their power. A possible method for direct conversion of their power into useful electrical power has been proposed[36]. In this scheme, shown in Figure 8, the positive ions are slowed down in the same way as in negative ion recovery. The current, $I_+$, is used to charge capacitors in a resonant modular convertor. These are then discharged, by switching on the transistors, through a transformer to produce useful electrical energy. A number of modules can be used in series and the electrical output connected in series or in parallel.

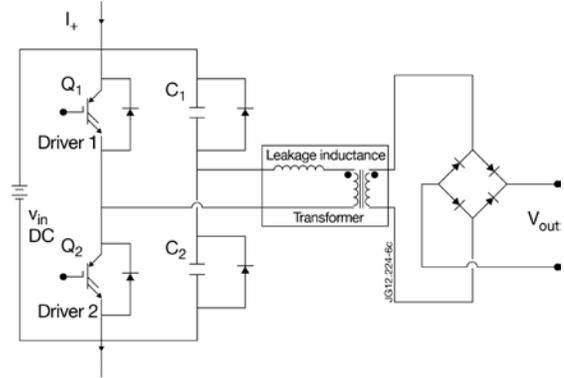

FIG. 8. Resonant modular convertor for direct conversion of positive ion current to useful power

Energy recovery systems have been developed for positive ion beamlines and tested for a negative ion beamline but no working system exists. There are challenges in the deceleration stages down to a few 10's of keV with minimal losses, the separation and collection of the residual ions, the beam halo may reduce the collection efficiency and it requires large plant and infrastructure.

## IV. SYSTEM EFFICIENCY STUDIES

In order to understand how the system efficiency is dependent on divergence, neutralisation efficiency and energy recovery a system code is under development[37]. The input parameters are given in Table 2 based on the ITER beamline. The transmission is calculated by the Beam Transmission and Re-ionisation (BTR) code and the laser power from equation 1. As designs progress the complexity and detail of the code will increase. The plasma neutraliser is not included but its efficiency between that for the gas and photo-detachment neutralisers. Figure 9 shows a comparison of the wallplug efficiency as the core divergence changes for a 15mrad halo containing 15% of the beam. Only with addition of energy recovery or conversion or higher neutralisation can the efficiency reach 40-45%. Using direct conversion of the positive ions improves the system efficiency slightly less than negative ion energy recovery due to the conversion efficiency to useful electrical power. The use of two technologies to reduce risk can also be considered as illustrated in Figure 10 which shows the dependence of wall plug efficiency for a 5mrad core and a 15mrad halo divergence beam for a gas neutraliser and a photo-neutraliser where both systems employ energy recovery methods. There relatively little gain in employing energy recovery with a photo-neutraliser operating at high neutralisation as there are few residual negative ions to make a large difference in the system efficiency.

TABLE 2. Parameters for system efficiency calculation.

| Ion source and beam | | Efficiencies and transmission | | Neutralisation and energy recovery | |
|---|---|---|---|---|---|
| Energy (MeV) | 1.0 | DC efficiency | 0.9 | **Gas neutraliser** | |
| $D^-$ current (A) | 59.1 | RF efficiency | 0.9 | Neutralisation efficiency | 0.58 |
| Electron/$D^-$ ratio | 1 | Stripping: No laser/laser | 0.29/0.24 | **Photo-detachment** | |
| Electron extraction voltage (kV) | 10 | Gas neutraliser efficiency | 0.58 | Neutralisation efficiency | 0.58-0.95 |
| Electron suppression voltage (V) | 15 | Core divergence (mrad) | 3-7 | Laser efficiency | 0.25 |
| Electron suppression current (A) | 166 | Halo divergence (mrad) | 15 | Laser power | Scaling formula |
| Filter field voltage (V) | 5 | Re-ionisation | BTR | **Energy recovery** | |
| Filter field current (A) | 6000 | Direct interception losses | BTR | Recovery energy (kV) | 25 |
| RF power (kW) | 800 | | | Recovery fraction | 0.8 |
| Incidentals (MW):  No Laser/Laser | 6/4.4 | | | Conversion efficiency for positive ions | 0.9 |

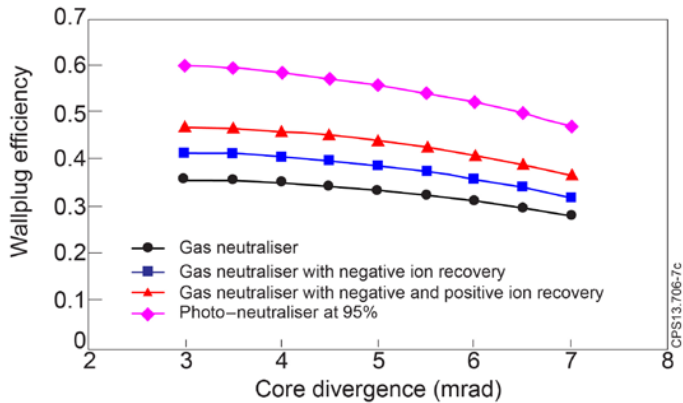

FIG. 9.(Color online) Wallplug efficiency dependence on core divergence

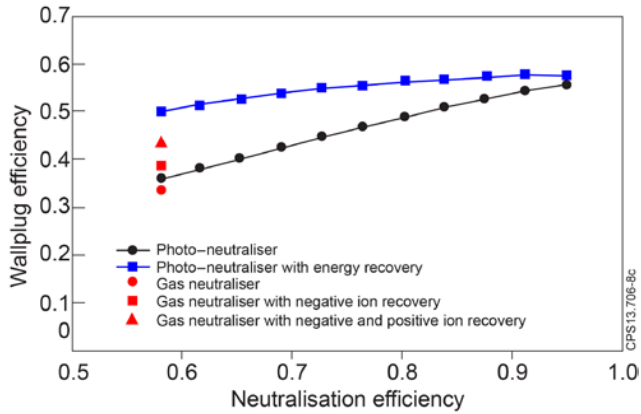

FIG. 10. (Color online) Wallplug efficiency dependence neutralisation efficiency under different scenarios

## ACKNOWLEDGEMENTS

This work was funded by the RCUK Energy Programme under grant EP/I501045 and the European Communities under the contract of Association between EURATOM and CCFE. To obtain further information on the data and models underlying this paper please contact PublicationsManager@ccfe.ac.uk. The views and opinions expressed herein do not necessarily reflect those of the European Commission. The author would like to thank Drs E Surrey, R Kemp, I Jenkins and D King for very useful discussions.